\documentclass[preprint,aps,prl,showpacs,preprintnumbers,amsmath,amssymb,superscriptaddress]{revtex4-1}

\usepackage{graphicx}
\usepackage{dcolumn}
\usepackage{bm}
\usepackage{bibtopic}
\usepackage{multirow}
\usepackage{subfigure}
\usepackage{xcolor}
\usepackage[normalem]{ulem}

\begin{document}


\title{Driving force for martensitic transformation in Ni$_{2}$Mn$_{1+x}$Sn$_{1-x}$}
  
      \author{Soumyadipta Pal}
      
      \affiliation{Department of Condensed Matter Physics and Material Sciences, S N Bose National Centre for Basic Sciences, Block-JD, Sector-III, Salt Lake, Kolkata - 700106, West Bengal, India.}
      \affiliation{Department of Physics, Calcutta Institute of Technology, Banitabla, Uluberia, Howrah - 711316, West Bengal, India.}
      
      \author{Sagar Sarkar}
      
      \affiliation{Department of Condensed Matter Physics and Material Sciences, S N Bose National Centre for Basic Sciences, Block-JD, Sector-III, Salt Lake, Kolkata - 700106, West Bengal, India.}
      
      \author{S K Pandey}
      
      \affiliation{Department of Condensed Matter Physics and Material Sciences, S N Bose National Centre for Basic Sciences, Block-JD, Sector-III, Salt Lake, Kolkata - 700106, West Bengal, India.}
      
      \author{Chhayabrita Maji}
      
      \affiliation{Department of Condensed Matter Physics and Material Sciences, S N Bose National Centre for Basic Sciences, Block-JD, Sector-III, Salt Lake, Kolkata - 700106, West Bengal, India.}
       \affiliation{Department of Materials Science, Indian Association for the Cultivation of Science, 2A \& 2B Raja S.C. Mullick Road, Jadavpur, Kolkata - 700032, West Bengal, India.}
       
      \author{Priya Mahadevan}
\email[Corresponding author: ] {priya.mahadevan@gmail.com  }
            
      \affiliation{Department of Condensed Matter Physics and Material Sciences, S N Bose National Centre for Basic Sciences, Block-JD, Sector-III, Salt Lake, Kolkata - 700106, West Bengal, India.}

\begin{abstract}  
The martensitic transformation in Ni$_{2}$Mn$_{1+x}$Sn$_{1-x}$ alloys has been investigated within 
\textit{ab-initio} density functional theory. The experimental trend of a martensitic transition happening beyond $x$ = 0.36 is captured within these calculations. The microscopic considerations leading to this are traced to increased Ni-Mn hybridization which results from the Ni atom experiencing a resultant force along a lattice parameter and moving towards the Mn atoms above a critical concentration. The presence of the lone pair electrons on Sn forces the movement of Ni atoms away from Sn. While band Jahn Teller effects have been associated with this transition, we show quantitatively that atleast in this class of compounds they have a minor role.   
\end{abstract}

\maketitle

\section{Introduction}
The interest in shape memory alloys has been driven by the enormous potential these materials represent in various fields ranging from medicine \cite{Morgan_MSEA_2004} to robotics \cite{Duerig_EASMAbook_1990} to aeronautics \cite{Hartl_Proced_2007}. This is because one has a diffusionless structural transition which involves the rearrangement of the position of the atoms in the solid. This process is entirely reversible and has been the driving force in using shape memory alloys in wide ranging applications. The microscopic mechanism that drives the transition is therefore of interest which would help us to identify which materials would undergo this martensitic structural transition (MST). Fermi surface nesting \cite{Opeil_PRL_2008} and soft phonon modes \cite{Wechsler_JMet_1953} have often been invoked to explain the MST, with the microscopic origin usually being associated with a band Jahn Teller effect \cite{Brown_JPCM_1999}. These reasons however do not seem to be valid across all systems. Examples among Heusler alloys are seen, where inspite of a soft phonon mode being found in the calculations, no MST has been observed \cite{Agduk_JALCOM_2012}. We consider the example of compounds given by Ni$_{2}$Mn$_{1+x}$Sn$_{1-x}$. The unusual feature of this class of compounds is that the martensitic transition is seen only for off-stoichiometric compositions where $x$ ranges from 0.36 to 0.80 in contrast to other martensites such as Ni$_{2}$MnGa where the transition is seen for stoichiometric members. The martensitic transition in Ni$_{2}$MnGa has been explained by Jahn-Teller effects \cite{Brown_JPCM_1999}. On the other hand the usual explanation offered in the case of Ni$_{2}$Mn$_{1+x}$Sn$_{1-x}$ is the increased hybridization between the Ni and Mn $d$ states being responsible for the observed martensitic transition \cite{Ye_PRL_2010}. This effect should be present in the stoichiometric composition also, and these ideas do not explain why one doesn't have a martensitic transition there.

In the present work we consider several compositions of Ni$_{2}$Mn$_{1+x}$Sn$_{1-x}$. Our calculations find a transition for $x$ = 0.375, 0.50, 0.625, 0.75 and 0.875, but find no transition for $x$ = 0.0 and 0.25, consistent with experiment. As our calculations are able to capture the experimental trend, we went on to examine the microscopic origin of the transition. Considering the $x$ = 0.50 composition, we have calculated the band structure and fit this to a microscopic tight binding model. This analysis was performed for the cubic structure as well as the martensitic structure and allows us to quantify the changes in the electronic structure. If the origin was Jahn-Teller distortions, one expects a change in the onsite energies as a result of the distortion. One however finds very small changes indicating that we must discard this model. 

Short Mn-Ni (Mn1-Ni) bonds equal to 2.624 \AA ~ are found in the parent compound Ni$_{2}$MnSn. When Mn replaces Sn atoms (referred to as Mn2) one has the same Mn2-Ni bondlength in the unrelaxed structure. However it is found that the system lowers its energy with Ni atoms moving towards the Mn2 atoms while there is a smaller decrease in the Ni-Mn1 bondlength. This seems surprising at first as all Mn-Sn bondlengths are the same to start with and so are the bondangles leading to similar matrix elements for the hopping between Mn and Ni. This is traced to the fact that while the exchange splitting on Mn1 and Ni are in the same direction, that on Mn2 is opposite. This results in a larger energy gain when Mn2 and Ni interact. At larger doping concentrations one gains energy from Mn1 and Ni interactions also. The increased Mn-Ni hybridization is not the driving force of the martensitic transition as, at $x$ = 0.25 one has shortened bonds, but no martensitic transition. As $x$ is increased, the interactions between Mn2 and the neighboring Mn1 leads to substantial deviation from a perfectly $d^{5}$ character on Mn1 that one finds otherwise. This then allows for energy gain from Mn1-Ni hybridization also. So a Ni atom moves towards both Mn1 and Mn2 and this is achieved by moving along the resultant force which is along a lattice parameter. Further energy lowering is possible by an elongation of that lattice parameters which results in the martensitic transition. The question that follows is why would the Ni atom move towards some of its neighbours over a centrosymmetric situation. This is traced to the presence of the lone pair on Sn. The system gains energy with Ni moving away from the Sn atoms resulting in reduced repulsion felt by the electrons on Ni from those on Sn. Hence a combination of Mn-Ni enhanced hybridization and reduced repulsion from the lone pair on Sn drive the martensitic transition. The lone pair electrons are not only present in Sn, but are also present in other elements like In and Sb. So, the mechanism proposed here should be valid for those systems also.   

\section{Methodology}                 
\textit{Ab-initio} electronic structure calculations are carried out using density functional theory (DFT) as implemented in the Vienna \textit{Ab-initio} Simulation Package (VASP) \cite{Kresse_PRB_1996}. We use the projected augmented wave implementation and work with the generalized gradient approximation (GGA) Perdew-Wang \cite{perdew_wang} for the exchange correlation functional. This has been seen to give a better description of the magnetism in Heusler compounds \cite{claudia_JPCM} and we also see 22-24\% enhanced magnetic moments in our GGA calculations compared to the LDA calculations. In off-stoichiometric Ni$_{2}$Mn$_{1+x}$Sn$_{1-x}$ ($x$ = 0.25, 0.375, 0.50, 0.625, 0.75, 0.875) the Mn at parent site is labelled Mn1 while the Mn doped at Sn site has been labelled Mn2. The magnetic interaction between Mn1 and Mn2 is antiferromagnetic, while that between Ni and Mn1 is ferromagnetic. Monkhorst-Pack k-points mesh of 10$\times$10$\times$10 for 16 atom unit cell and 10$\times$5$\times$10 for 32 atom supercell were used to perform the k space integrations and a 
cut-off energy of 340~eV was used to determine the plane-waves used in the basis. The lattice parameters of the unit cell as well as the ionic positions have been optimised in each of the cases considered and the optimised value
is mentioned in the text. An analysis of the electronic structure has been carried out in terms of the band dispersions as well as the partial density of states (PDOS) calculated using spheres of radii $\sim$ 1.3 \AA ~ around each atom. The magnetic moments have been reported within the same spheres. Additionally an analysis of the electronic structure has been carried out using an interface of VASP to WANNIER90 \cite{Mostofi_CPC_2008,Marzari_PRB_1997,Souza_PRB_2001}. 
A basis consisting of Ni $s$ and $d$, Mn $s$ and $d$ as well as Sn $p$ states.
A mapping of the Bloch states is made onto Wannier functions,
localized on the respective atoms with their angular parts given by the relevant spherical harmonics, via
a unitary transformation. A unique transformation is obtained with the requirement of minimizing the 
quadratic spread of the Wannier functions. The criterion of convergence was that the spread
changed by less than 10$^{-6}$ between successive iterations. Once the transformation
matrices are determined, one has a tight binding representation of the Hamiltonian in the basis of the
maximally localized Wannier functions.
 
\section{Results and Discussions}
\begin{figure}[h]
\includegraphics[angle=0,width=11cm,height=8cm]{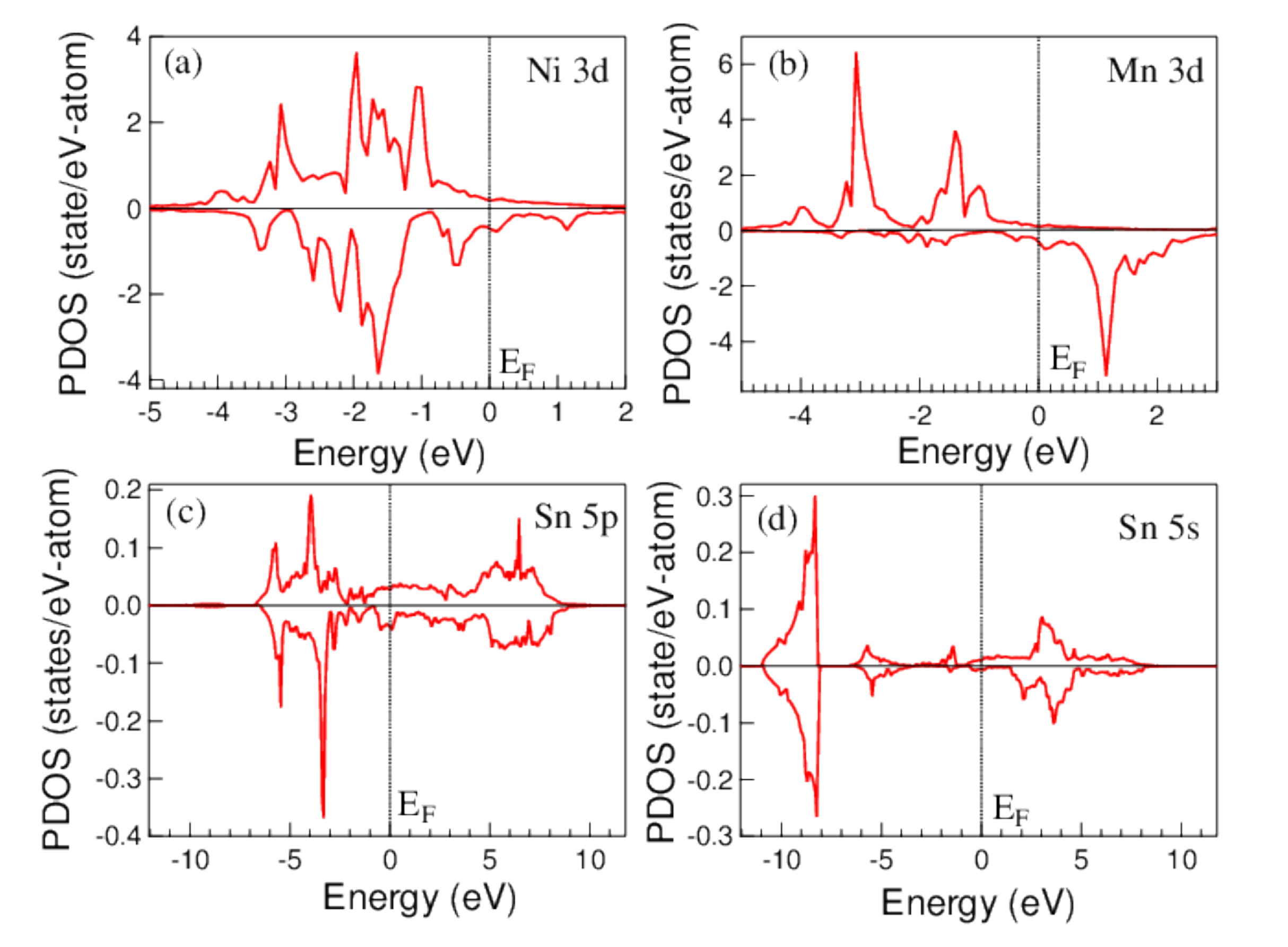}
\caption{(Color online) Atom and angular momentum projected partial density of states for (a) Ni 3$d$, (b)Mn 3$d$, (c) Sn 5$p$ and (d) Sn 5$s$ in the parent compound Ni$_{2}$MnSn. The Fermi energy E$_{F}$ is at 0 eV.}
\label{fig1}
\end{figure}

The experimentally reported unit cell of Ni$_{2}$MnSn is cubic with a lattice parameter of 6.05 \AA \cite{Krenke_PRB_2005}. Carrying out an optimization of the lattice parameter within our calculations, the theoretical lattice parameter is found to be 6.06 \AA ~ and the structure remains cubic. The distances between Ni and Mn as well as between Ni and Sn are found to be 2.62 \AA. In contrast to what we find in Ni$_{2}$MnSn, Ni$_{2}$MnGa is found to undergo a structural transformation and favors a tetragonal phase at low temperature. This suggests that an important role is played by the p-element in inducing the martensitic transformation. In order to understand this further we examine the atom and orbital projected partial density of states. This is shown in Fig. \ref{fig1}. 
Ni $d$ states are found to contribute in the energy window -4 to 2 eV, which is also the energy window in which Mn 3$d$ states contribute. Sn 5$p$ states are found to have a low weight in the same energy window. Sn 5$s$ states are localized and are found to contribute -8 to -10 eV below the Fermi level, with low weight in the unoccupied part. These states have been referred to as lone pair states as they are not involved in any bonding with the neighboring atoms. This indicates that the electronic structure of Ni$_{2}$MnSn emerges from the bonding of the Ni 3$d$ and Mn 3$d$ states. Examining the Mn $d$ PDOS, one finds that
the majority spin states are completely filled while the minority spin states 
are empty, indicating a $d^{5}$ configuration at the Mn site. The 
exchange splitting of the Mn $d$ states and the Ni $d$ states are found to be
in the same direction.

Now when one of the Sn atoms is replaced by Mn, corresponding to the composition Ni$_{2}$Mn$_{1.25}$Sn$_{0.75}$, we find that despite allowing for changes in the cell shape upon optimization, the structure remains cubic. The optimized lattice constant is found to 5.99 \AA ~ as against 6.06 \AA ~ that was found for the parent compound. The absence of a tetragonal transition at this composition, which is usually associated with the existence of a martensitic transition, is consistent with experiment. There is also a volume contraction found when we replace Sn with Mn. This is expected as the ionic radius of Mn is smaller than that of Sn. The bond lengths between Ni-Mn1, Ni-Mn2 and Ni-Sn are found to be 2.57 \AA ~, 2.52 \AA ~ and 2.62 \AA ~ as against the bondlengths of 2.59 \AA ~ found before the atomic relaxations. There is a reduction in the Ni-Mn bondlengths, with a larger reduction in the Ni-Mn2 bondlengths.

We then consider the composition $x$ = 0.375 in the formula Ni$_{2}$Mn$_{1+x}$Sn$_{1-x}$. This is close to the composition $x$ = 0.36 at which point one finds the onset of the martensitic transitions in experiment. This can be realized in a 32 atom supercell. Starting with a cubic unit cell one finds that a tetragonal unit cell is favored at the end of the unit cell optimization with lattice parameters $a =$ 6.83 \AA, $b \approx c$ = 5.55 \AA , resulting in a tetragonality $a/c \approx$ 1.23. However, the tetragonality reported in experiment \cite{Krenke_PRB_2005,Sutou_APL_2004,Brown_JPCM_2006} is $\sim$ 1.10. This might be due to structural disorder which are not accounted for in the theoretical calculation. The difference in the lattice parameters by GGA based DFT and the experiment has been also reported for Ni-Mn-Ga system \cite{Barman_PRB_2007}. Moreover, one finds a reduction in some of the Ni-Mn bond lengths. These are found to be in the range 2.50-2.64 \AA ~ for Ni-Mn1, 2.53-2.54 \AA ~ for Ni-Mn2. Additionally there is a substantial increase of the Ni-Sn bond lengths from the stoichiometric compound. The question that follows is whether this aids the tetragonality and how. Understanding this would help us to explain the observed martensitic phase transition.

\begin{figure}[ht]
\includegraphics[angle=0,width=16cm,height=13cm]{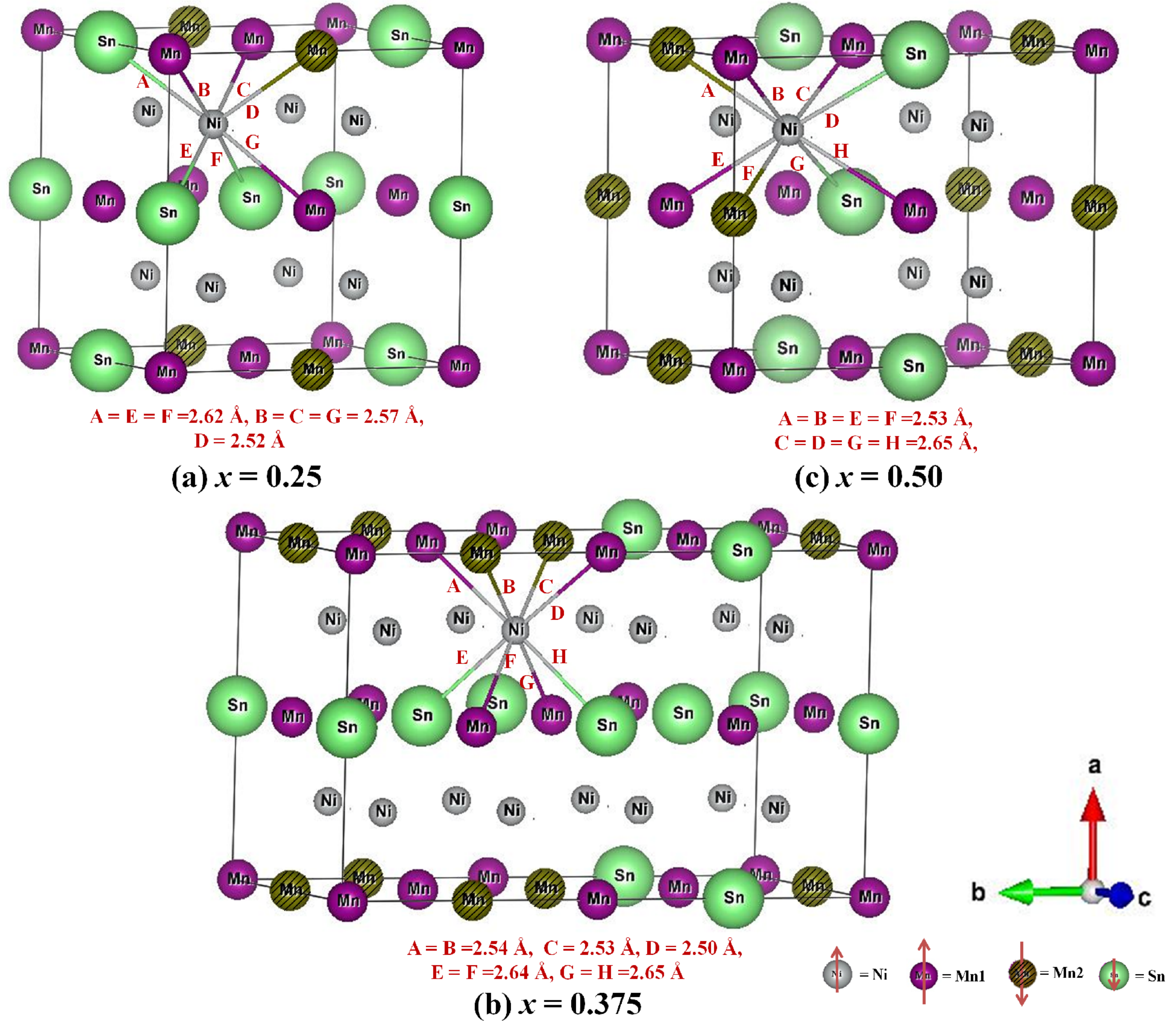}
\caption{(Color online) Ni-Mn1 and Ni-Mn2 bond distances in \AA ~ unit, after full relaxation of cubic structure of composition $x$ = 0.25, 0.375 and 0.50 of Ni$_{2}$Mn$_{1+x}$Sn$_{1-x}$. The direction of magnetic moment on each atom has been indicated by arrow.}
\label{fig2}
\end{figure}

We then continue the discussion by considering the composition $x$ = 0.50. 
Here again one finds that there is a reduction in Ni-Mn1 and Ni-Mn2 bond lengths. Additionally there is a tetragonal unit cell which is found to be favored for different combinations of $x$ = 0.50 which is the indicator of the martensitic transition. 

In Fig. \ref{fig2}, fully relaxed crystal structures of initial cubic structure of composition $x$ = 0.25, 0.375 and 0.50 of Ni$_{2}$Mn$_{1+x}$Sn$_{1-x}$ have been shown. 
In each case the optimized Ni-Mn1 and Ni-Mn2 bond lengths have been indicated. 
Examining the structures closely, we can identify a pattern that emerges in the relaxations. 
Ni atoms are found to move towards Mn2 atoms, with their bondlengths with Sn
increasing. Their movement towards Mn1 atoms is less though 
at larger concentrations they move towards those Mn1 atoms which are 
nearest neighbors of Mn2. This is consistent with extended x-ray absorption fine structure (EXAFS) measurements of Ni$_{2}$Mn$_{1.4}$Sn$_{0.6}$ \cite{Bhobe_JPCM_2008} where one finds that Ni-Mn bond lengths decrease upon martensitic transformation. In the low temperature martensitic phase in the case of Ni K-edge, Ni-Mn and Ni-Sn bond lengths are 2.569 \AA ~ and 2.607 \AA ~ changing from 2.595 \AA ~.
This seems surprising at first and the question
we ask next is what are the underlying energetics that dictate the Ni movement.
At $x$ = 0.25 we find that both Mn1 and Mn2 have an almost $d^{5}$ configuration
though their exchange splittings are opposite in direction. This results
in stronger hybridization between Ni and Mn2 atoms, compared to that with Mn1.
This is the reason the Ni atom moves towards Mn2. As $x$ is increased, one has
Mn1 and Mn2 atoms occupying nearest neighbor sites. They interact with 
each other and as a result one finds a deviation of Mn1 from a $d^{5}$ configuration found at $x$ = 0.25. This results in a smaller moment on those Mn1 atoms which are closer to Mn2 and is found to be 3.383 $\mu_{B}$ as against 3.486 $\mu_{B}$ for the stoichiometric Ni$_{2}$MnSn. 
Now one finds that Ni atoms can gain energy from interactions
with both Mn1 (only the low moment atoms) and Mn2. This is what we find for
$x$ = 0.375 and beyond. The Ni atoms move towards both Mn1 and Mn2. The resultant 
movement is along a lattice parameter. Hence in each case one finds that the
lattice parameter increases along the direction of movement
and there is further lowering of energy. The natural question that follows 
is why does the Ni atom move away from a centro-symmetric position
towards the Mn atoms. 

\begin{figure}[htb]
\includegraphics[angle=0,width=11cm,height=7cm]{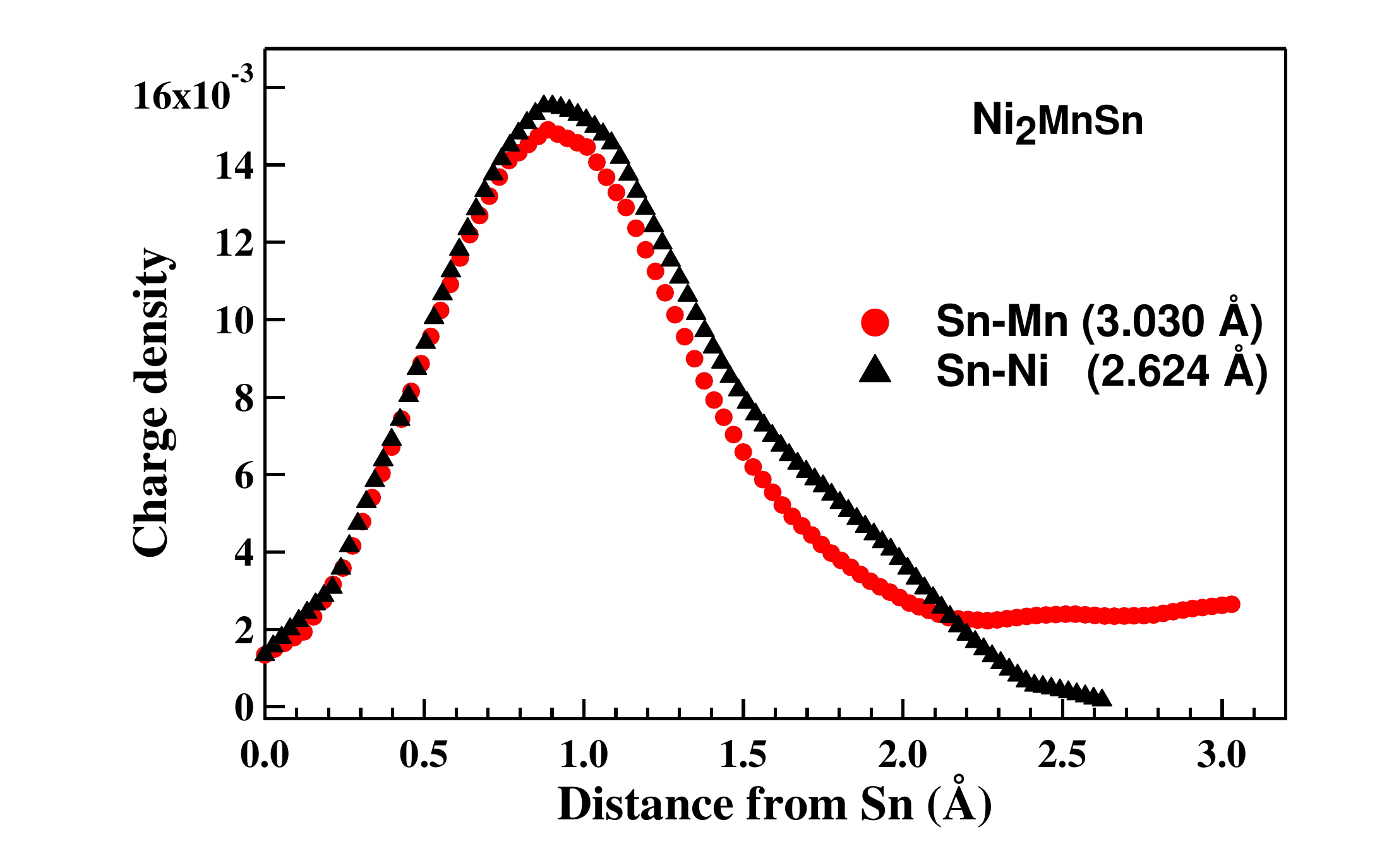}
\caption{(Color online) Line charge density of the Sn 5$s$ states along the Sn-Mn and Sn-Ni bond. The respective bondlengths are indicated in parentheses.}
\label{fig3}
\end{figure}

In each of the structures shown in Fig. \ref{fig2}, one
finds that the movement of the Ni atoms 
is away from the Sn atoms. The reason for this is
apparent when we plot the line profile of the Sn 5$s$ charge density along the
Sn-Mn and Sn-Ni bonds of Ni$_{2}$MnSn in Fig \ref{fig3}. 
The chosen Mn atom is further away and is at 3.03 \AA ~ from the Sn atom. However, the Ni atom is at 2.62 \AA ~ from the Sn atom. If there were strong covalent interactions between the Ni and Sn atoms, one would
expect the charge density to be more delocalized along the bond.
Instead we find that the charge density is more localized along the Sn-Ni
bond than along the Sn-Mn bond, the spread reflecting the increased
separation of the pair of atoms. This verifies that the driving force of the distortions is additionally the repulsion the electrons on Ni face from those
on Sn.

\begin{figure}[t]
\includegraphics[angle=0,width=8cm,height=12cm]{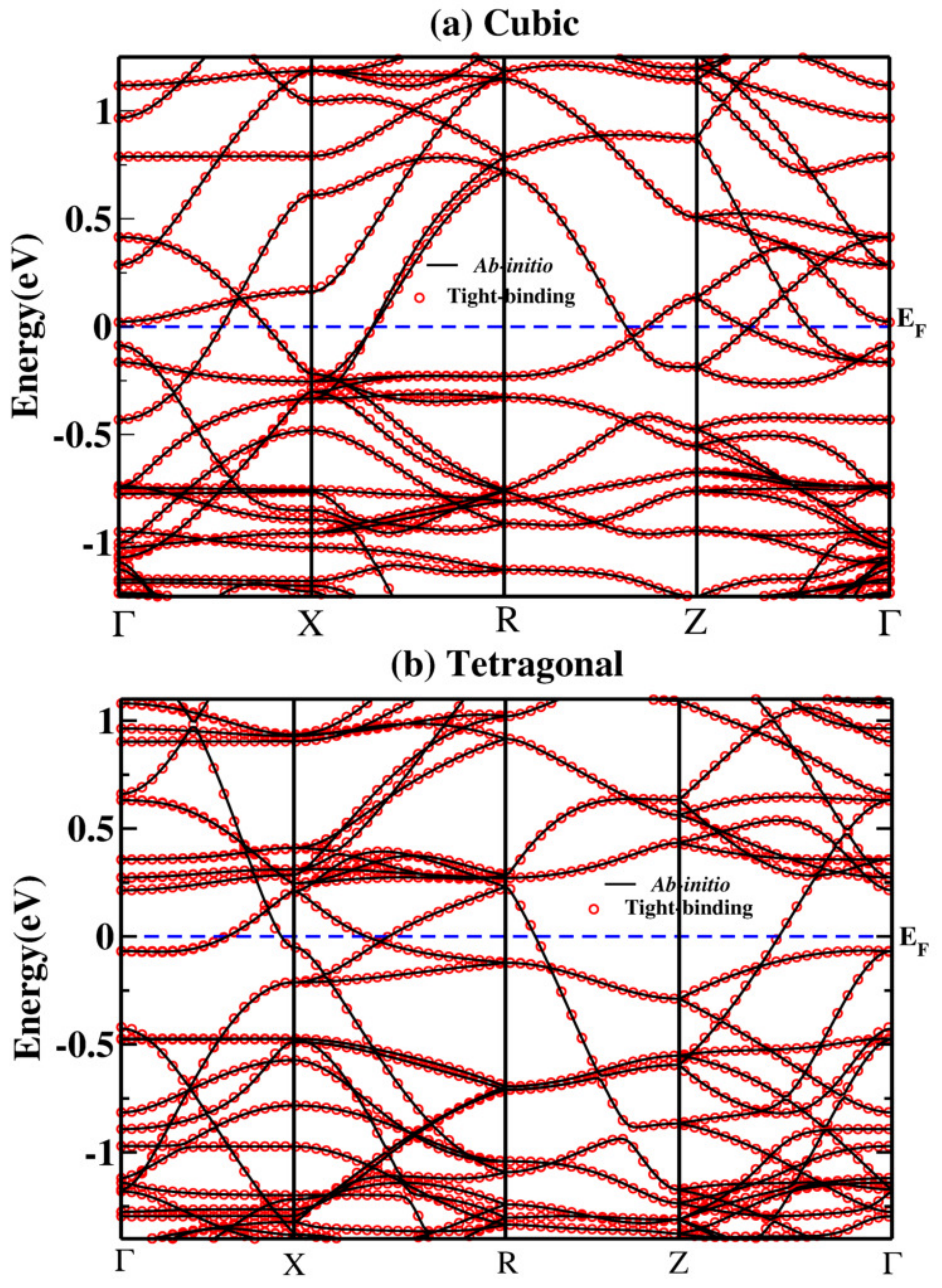}
\caption{(Color online) The tight-binding (circles) as well as the \textit{ab-initio} band structure (solid line) of Ni$_{2}$Mn$_{1.5}$Sn$_{0.5}$ 
in the (a) cubic and the (b) fully relaxed tetragonal phase.}
\label{fig4}
\end{figure}

An alternate explanation for the martensitic transition that has been offered 
has been the band Jahn Teller effect.
The idea is that Jahn Teller distortions lift the degeneracy of the $d$ 
orbitals. This is aided by tetragonality and hence the conclusion that Jahn-Teller distortions drive the martensitic transition.
In order to quantify this we map the \textit{ab-initio} band dispersions for $x$ = 0.50 to a tight binding model which included $s$ and $d$ orbital states of Ni and Mn and $p$ orbital states of Sn in the basis. Maximally localized wannier functions were considered for the radial parts of the wavefunction. The tight-binding bands are superposed on the \textit{ab-initio} bands of unstable cubic and fully relaxed tetragonal structure in Fig. \ref{fig4} (a) and (b), respectively. 

In both cases, one finds an excellent mapping of the \textit{ab-initio} band structure within the tight-binding model. The relative on-site energy of Ni d states with respect to $d_{\text{xz}}$ spin up state are listed in Table \ref{Table1}. The changes one finds in the energies are small. This suggests that Jahn-Teller distortions cannot be the driving force for the martensitic transitions seen in this system. 
\begin{table}[htb]
\centering
\small
\caption{Relative on-site energies with respect to spin up $d_{\text{xz}}$ state of Ni in cubic and fully relaxed tetragonal structure of of Ni$_{2}$Mn$_{1.5}$Sn$_{0.5}$}
\resizebox{10cm}{!}{
\begin{tabular}{cccccccc}
\hline
& & & ~~~~Cubic Unit cell & & &~~~~~Tetragonal unit cell  &\\ 
Atom&&Orbital& Spin up & Spin down& &Spin up & Spin down \\
\hline
&&$d_{\text{z}^{2}}$ & 0.11 & 0.27& & -0.01 & 0.24 \\
& &$d_{\text{xz}}$ & 0.0 & 0.12 & &0.0 & 0.05\\
Ni&&$d_{\text{yz}}$ & 0.07 & 0.23 & &0.06 & 0.03 \\
&&$d_{\text{x}^{2}-\text{y}^{2}}$ & 0.10 & 0.30 & &0.02 & 0.10\\
&&$d_{\text{xy}}$ &0.03 &0.15 & &0.03 &0.19\\
\hline
\end{tabular}}
\label{Table1}
\end{table} 
It is important to note that in the spin down channel reduced exchange splitting is possibly due to transfer of charge to Ni atoms. This might be from Sn $s$ states. Thus our analysis reveals that the lone pair effect of Sn 5$s$ electrons is the main triggering factor for additional tetragonality.

\section{Conclusions}
In summary, we have investigated the structural properties of Ni$_{2}$Mn$_{1+x}$Sn$_{1-x}$ by means of \textit{ab-initio} density functional theory. We have obtained martensitic transformation for $x \geq$ 0.375 which is in good agreement with the experimental value of $x \geq$ 0.36. The relative on-site energies of Ni $d$ states in cubic and fully relaxed tetragonal structure of Ni$_{2}$Mn$_{1.5}$Sn$_{0.5}$ reveals that the changes in the energies are very small and that the Jahn-Teller effect cannot be the driving force for the martensitic transitions seen in this system. It is the Ni-Mn hybridization and Sn lone pair effect on Ni which makes the cubic structure unstable and triggers the structural transformation for Ni-Mn-Sn systems. The microscopic considerations that result in the mechanism becoming operative in non-stoichiometric compositions are discussed.

We acknowledge funding from DST Nanomission under the umbrella of the Thematic Unit of Computational Material Science and DAE.
S.S acknowledges CSIR India for his fellowship.


\begin{thebibliography}{100}

\bibitem{Morgan_MSEA_2004}
N.B. Morgan, Mat. Sci. Eng. A \textbf{378}, 16 (2004).

\bibitem{Duerig_EASMAbook_1990}
T. W. Duerig, K. N. Melton, D. St$\ddot{o}$ckel, and C. M. Wayman, Engineering Aspects of Shape Memory Alloys, Butterworth-Heinemann Ltd. (1990).

\bibitem{Hartl_Proced_2007}
D J Hartl and D C Lagoudas, Proc. IMechE Part G: J. Aerospace Engineering. \textbf{221}, 535 (2007).

\bibitem{Opeil_PRL_2008}
C. P. Opeil, B. Mihaila, R. K. Schulze, L. Ma$\tilde{n}$osa, A. Planes, W. L. Hults, R. A. Fisher, P. S. Riseborough, P. B. Littlewood, J. L. Smith, and J. C. Lashley, Phys. Rev. Lett. \textbf{100}, 165703 (2008).

\bibitem{Wechsler_JMet_1953}
M. S. Wechsler, D. S. Lieberman, and T. A. Read, J. Met. \textbf{5}, 1503 (1953).

\bibitem{Brown_JPCM_1999}
P. J. Brown, A. Y. Bargawi, J. Crangle, K. -U. Neumann, and K. R. A. Ziebeck, J. Phys.: Condens. Matter \textbf{11}, 4715 (1999).

\bibitem{Agduk_JALCOM_2012}
S. A$\breve{g}$duk and G. G$\ddot{o}$ko$\breve{g}$lu, J. Alloys Compd. \textbf{511}, 9 (2012).

\bibitem{Ye_PRL_2010}
M. Ye, A. Kimura, Y. Miura, M. Shirai, Y. T. Cui, K. Shimada, H. Namatame, M. Taniguchi, S. Ueda, K. Kobayashi, R. Kainuma, T. Shishido, K. Fukushima, and T. Kanomata, Phys. Rev. Lett. \textbf{104}, 176401 (2010).

\bibitem{Kresse_PRB_1996}
G. Kresse and J. Furthm$\ddot{\text{u}}$ller, Phys. Rev. B \textbf{54}, 11169 (1996).

\bibitem{perdew_wang}
J.P. Perdew, K. Burke and Y. Wang, Phys. Rev. B \textbf{54}, 16533 (1996).

\bibitem{claudia_JPCM}
H. C. Kandpal, G.H. Fecher and Claudia Felser, J Phys. D \textbf{40}, 1507 (2007).

\bibitem{Mostofi_CPC_2008}
A. A. Mostofi, J. R. Yates, Y.-S. Lee, I. Souza, D. Vanderbilt, and N. Marzari, Comput. Phys. Commun. \textbf{178}, 685 (2008).

\bibitem{Marzari_PRB_1997}
N. Marzari and D. Vanderbilt, Phys. Rev. B \textbf{56}, 12847 (1997).

\bibitem{Souza_PRB_2001}
I. Souza, N. Marzari and D. Vanderbilt, Phys. Rev. B \textbf{65}, 035109 (2001).

\bibitem{Krenke_PRB_2005}
T. Krenke, M. Acet, E. F. Wassermann, X. Moya, L. Ma$\tilde{\text{n}}$osa, and A. Planes, Phys. Rev. B \textbf{72}, 014412 (2005).

\bibitem{Sutou_APL_2004}
Y. Sutou, Y. Imano, N. Koeda, T. Omori, R. Kainuma, K. Ishida, and K. Oikawa, Appl. Phys. Lett. \textbf{85}, 4358 (2004).

\bibitem{Brown_JPCM_2006}
P. J. Brown, A. P. Gandy, K. Ishida, R. Kainuma, T. Kanomata, K-U Neumann, K. Oikawa, B. Ouladdiaf, and K .R. A. Ziebeck, J. Phys.: Condens. Matter \textbf{18}, 2249 (2006).

\bibitem{Barman_PRB_2007}
S. Banik, R. Ranjan, A. Chakrabarti, S. Bhardwaj, N. P. Lalla, A. M. Awasthi, V. Sathe, D. M. Phase,
P. K. Mukhopadhyay, D. Pandey, and S. R. Barman, Phys. Rev. B \textbf{75}, 104107 (2007).

\bibitem{Bhobe_JPCM_2008}
P A Bhobe, K R Priolkar, and P R Sarode, J. Phys.: Condens. Matter \textbf{20}, 015219 (2008).


\end{thebibliography}
\end{document}